\documentclass[prerpint,floats,aps,epsfig]{revtex4}
\usepackage{epsfig}
\usepackage{amsmath}
\usepackage{amsfonts}
\def\({\left(}
\def\){\right)}
\def\[{\left[}
\def\]{\right]}
\def\non{ \nonumber }
\def\b{\beta}

\def\s{\sigma}

\def\be{\begin{equation}}
\def\ee{\end{equation}}
\def\bdm{\begin{displaymath}}
\def\edm{\end{displaymath}}
\def\bea{\begin{eqnarray}}
\def\eea{\end{eqnarray}}

\def\s{\sigma}

\newcommand{\p}{\partial}

\newcommand{\la}{\langle}
\newcommand{\ra}{\rangle}
\newcommand{\rd}{\mbox{d}}
\newcommand{\ri}{\mbox{i}}
\newcommand{\re}{\mbox{e}}

\begin{document}
\draft
\title{A Model with Propagating  Spinons beyond  One Dimension. } 
 \author{F. A. Smirnov  and  A. M. Tsvelik$^*$ }
\affiliation{LPTHE, Tour 16, 1-er \'etage, 4, pl. Jussieu 75252, Paris Cedex 05, France\\
$^*$Department of  Physics, Brookhaven 
National Laboratory, Upton, NY 11973-5000, USA}
\date{\today}
\begin{abstract}
For the model of frustrated spin-1/2 Heisenberg magnet described in A. A. Nersesyan and A. M. Tsvelik, (Phys. Rev. B{\bf 67}, 024422 (2003))  we calculate correlation functions of staggered magnetization and dimerization. The model is formulated as a collection of antiferromagnetic chains weakly coupled by a frustrated exchange interaction. The calculation done for the case  of four chains 
demonstrates that these  functions do not vanish. Since the correlation functions in question factorize into a product of correlation functions of spinon creation and annihilation operators, this  constitutes  a proof that spinons in this model  propagate in the direction perpendicular to the chains.   
\end{abstract}
\maketitle
\section{Introduction}

In \cite{NTsv} Alexander Nersesyan and one of the authors presented a proof  of existence of fractional quantum number excitations  in a model describing  a certain frustrated magnet in the number of dimensions greater than one. This  
magnet consists of  spin-1/2 antiferromagnetic Heisenberg chains weakly coupled by a frustrated antiferromagnetic interaction:
\bea
H = \sum_{j,n}\left\{J_{\parallel}{\bf S}_{j,n} \cdot {\bf S}_{j+1,n} + 
\sum_{\mu = \pm 1} \left[ J_r(n,n + \mu) {\bf S}_{j,n} + J_d(n,n + \mu)
\left( {\bf S}_{j+1,n} + {\bf S}_{j-1,n} \right) \right] \cdot {\bf S}_{j,n + \mu} \right\},
\eea
 where ${\bf S}_{j,n}$ are spin-1/2 operators, and 
$J_{\parallel} >>  J_r , J_d > 0$.
The interaction pattern of this model resembles  the flag of American Confederation which gave the model its name.  Fractional quantum number excitations appear  when $J_r(n,n + \mu) = 2J_d(n,n + \mu)$. If this condition is fulfilled,   the interaction between  staggered components of the magnetization on chains $n$ and $n +\mu$ vanishes. The weakness of the interchain coupling allows us to employ the continuous description. In the  continuum limit each Heisenberg chain is represented  by the SU$_1$(2) Wess-Zumino-Novikov-Witten (WZNW) model and the {\it relevant} part of the interchain interaction is reduced to the interaction of  spin currents with different chirality:
\bea
H = \sum_{n=1}^N \left[H_n + \gamma(n,n +\mu)\int \rd x({\bf J}_n\bar{\bf J}_{n + \mu} + {\bf J}_{n + \mu}\bar{\bf J}_{n})\right] \label{model}
\eea
Here  $H_n$ is the SU$_1$(2) WZNW Hamiltonian
\bea
H_{n} = \frac{2\pi v}{3}\int \rd x \left(:{\bf J}_n^2: + :{\bar{\bf J}_n}^2:\right)\label{WZNW}
\eea
where $v = \pi J_{\parallel} a_0/2$ is the spin velocity,  $J_n^a(x), \bar J_n^a(x)$ are operators representing holomorphic and antiholomorphic currents belonging to  the SU(2)$_1$ Kac-Moody algebra. The currents with different $n$ commute. The coupling constant $\gamma(n,n +\mu) \sim J_r(n,n +\mu)$. Continuum limit Hamiltonian (\ref{model}) coincides with the one introduced by Emery, Kivelson and Zachar\cite{EmKiv} in the context of theory of stripes. We emphasise that the interchain interaction in this frustrated model remains relevant and generates spectral gaps. Therefore this model does not belong to the class of  sliding Luttinger liquid models  where soliton excitations remain confined to the chains. 

Though  the authors of \cite{NTsv} have managed to proof the  existence of spin $S=1/2$ excitations in model (\ref{model}),  it remained unclear whether these excitations are able to  propagate in the direction transverse to the chains or remain confined. In the Confederate Flag model the staggered magnetization operator creates two non-interacting spinons (the absence of interaction follows from the fact that these spinons belong to different sectors of the Hamiltonian $H^{+}, H^{-}$, -see below). Therefore one way to resolve the problem of propagation  is to calculate correlation functions 
of the staggered magnetizations between  different  chains. If these correlation functions do not vanish, the excitations propagate. The difficulty is that such correlation functions are essentially non-perturbative objects (they remain  zero in any order of perturbation theory) and to calculate them one has to somehow go beyond perturbation theory. The corresponding methods are available for a finite number of chains (namely $N =2,3,4$ where exact solutions are available).  Though the most interesting case corresponds to an infinite number of chains $N \rightarrow \infty$,  information obtained  for finite $N$ may also provide valuable insights. The  calculations of the interchain correlation function of staggered magnetizations done in \cite{NTsv} for the case of two chains gave  a nonvanishing  answer, but one may argue that $N =2$ is too small a number to allow even a qualitative extrapolation to the thermodynamic limit. Similar calculations done for $N =4$ could not distinguish between intra and interchain correlation functions.  In this paper we perform accurate  calculations for  the case of four chains $N = 4$. We find that the interchain 
correlation functions do not vanish. 

 In  the continuum  limit the staggered magnetization ${\bf N} = (-1)^j{\bf S}_j$ and the dimerization $\epsilon = (-1)^j({\bf S}_j{\bf S}_{j +1})$  operators become smooth fields. For the chain number $n$ they are expressed through  matrix elements of the $S =1/2$ primary field of the $n$-th WZNW model:
\bea
\hat g_{\s\s'}(n;x) = \delta_{\s\s'}\epsilon(n;x) + \ri(\vec\sigma)_{\s\s'}{\bf N}(n;x) \label{g}
\eea
The action for model (\ref{model}) is local in $g$ and posseses the global  SU$_R$(2)$\times$SU$_L$(2) symmetry. Namely, the action remains invariant under the following transformations:
\bea
&& g_{2n} \rightarrow g_{2n}V, ~~ g_{2m+1} \rightarrow V^+g_{2m+1}, \nonumber\\
&& g_{2n} \rightarrow Ug_{2n}, ~~ g_{2m+1} \rightarrow g_{2m+1}U^+,
\eea
where $V,U$ are coordinate-independent SU(2) matrices. This symmetry dictates the following form of the two-point correlation functions:
\bea
&& \la\la g_{\s_1\s_2}(\tau,x;2n) g^+_{\s_3\s_4}(0,0;2m)\ra\ra = \delta_{s_1\s_4}\delta_{s_2\s_3}{\cal D}_{2n,2m}(\tau,x)\nonumber\\
&& \la\la g_{\s_1\s_2}(\tau,x;2n) g_{\s_3\s_4}(0,0;2m +1)\ra\ra = \delta_{s_1\s_4}\delta_{s_2\s_3}{\cal D}_{2n,2m+1}(\tau,x)
 \eea
where ${\cal D}$ do not contain spin indices. 
Substituting Eq.(\ref{g}) in the above equations we obtain the following important relation between correlation functions of staggered energy density and staggered magnetization:
\bea
&& \la\la\epsilon_{2n}(\tau,x)\epsilon_{2m}(0,0)\ra\ra =  \la\la N^a_{2n}(\tau,x)N^a_{2m}(0,0)\ra\ra \nonumber\\
&& \la\la\epsilon_{2n +1}(\tau,x)\epsilon_{2m +1}(0,0)\ra\ra =  \la\la N^a_{2n +1}(\tau,x)N^a_{2m +1}(0,0)\ra\ra \nonumber\\
 &&\la\la\epsilon_{2n}(\tau,x)\epsilon_{2m +1}(0,0)\ra\ra =  - \la\la N^a_{2n}(\tau,x)N^a_{2m +1}(0,0)\ra\ra \nonumber\\
\eea
 Now let us recall the important feature of Hamiltonian (\ref{model}): as was noticed in \cite{NTsv}, it separates into a sum of two commuting parts: $H = H^+ + H^-$ (such structure of the Hamiltonian was first observed  in the  context of Kondo lattice in \cite{orignac}) . Therefore eigenstates are separated into two sectors with different parity. The Hamiltonian density in the plus parity sector is 
\bea
&&{\cal H}^+ = \frac{2\pi v}{3}\left(:{\bf J_1}^2: + :{\bf J_3}^2:+ 
:{\bar{\bf J}_2}^2: + :{\bar{\bf J}_4}^2:\right) + (\lambda {\bf J}_1 + \lambda' {\bf J}_3)(\lambda \bar{\bf J}_2 + 
\lambda'\bar{\bf J}_4) \label{model2}
\eea
For the purposes of  this paper we find it convenient to consider different copling constants $\lambda \neq \lambda'$ such that the difference between these coupling constants is small ($|\lambda -\lambda'| << \lambda$). Physically this means that we still consider the frustrated interchain interactions, but allow their amplitudes to vary between different chains.  

 By using the Abelian bosonization procedure for the SU$_1$(2) currents one can write the Hamiltonians $H^{\pm}$  in terms of bosonic fields $\varphi, \bar\varphi$ living on each chain (see \cite{NTsv} for the details). In this representation fields $\varphi_{1,3}$ ($\bar\varphi_{1,3}$) interact with the fields
$\bar\varphi_{2,4}$ ($\varphi_{2,4}$) such that  
\bea
H = H^+[\varphi_{1,3};\bar\varphi_{2,4}] + H^-[\bar\varphi_{1,3};\varphi_{2,4}]\label{plusm}
\eea 
 The convenience of this representation is related to the fact that  it  allows to represent $\hat g(j)$ in a factorized form:
\bea
&&\hat g_{\s\s'} = \frac{1}{\sqrt 2}C_{\s\s'}:\exp[-\ri\sqrt{2\pi}
(\s\varphi + \s'\bar\varphi)]: \equiv C_{\s\s'}z_{\s}\bar z_{\s'},\nonumber\\
&&C_{\s\s'} = \re^{\ri\pi(1 - \s\s')/4}, \label{factor}
\eea
where 
\bea
 z_{\s} = \exp[\ri\s\sqrt{2\pi}\varphi], ~~ \bar z_{\s} = \exp[-\ri\s\sqrt{2\pi}
\bar\varphi],  ~~(\s = \pm 1).
\eea
Eqs.(\ref{plusm},\ref{factor}) select a  set of potentially non-vanishing interchain correlators:
\bea
&&G_{13} = \la\la \re^{\ri\sqrt{2\pi}\varphi(1)}\re^{-\ri\sqrt{2\pi}\varphi(3)} \ra\ra, ~~ \bar G_{13}= \la\la \re^{\ri\sqrt{2\pi}\bar\varphi(1)}\re^{-\ri\sqrt{2\pi}\bar\varphi(3)}\ra\ra\nonumber\\
 &&D_{12} = \la\la \re^{\ri\sqrt{2\pi}\varphi(1)}\re^{\ri\sqrt{2\pi}\bar\varphi(2)} \ra\ra, ~~ \bar D_{12} = \la\la \re^{\ri\sqrt{2\pi}\bar\varphi(1)}\re^{-\ri\sqrt{2\pi}\varphi(2)}\ra\ra
\eea
From these correlators one can obtain correlation functions of both vector ${\bf N}$ and scalar $\epsilon$ staggered fields:
\bea
 &&\la\la\epsilon_{2n}(\tau,x)\epsilon_{2m}(0,0)\ra\ra =  \la\la N^a_{2n}(\tau,x)N^a_{2m}(0,0)\ra\ra  = |G_{2n,2m}(\tau,x)|^2\nonumber\\
 &&\la\la\epsilon_{2n}(\tau,x)\epsilon_{2m +1}(0,0)\ra\ra =  - \la\la N^a_{2n}(\tau,x)N^a_{2m +1}(0,0)\ra\ra = |D_{2n,2m+1}(\tau,x)|^2 \label{corr}
\eea 
 According to these formulas  spins on neighboring chains are oriented antiferromagnetically. We emphasise that this conclusion follows just from the symmetry arguments and the chiral decoupling of the operators (\ref{factor}) and the Hamiltonian (\ref{plusm}). 

\section{Exact solution of model (\ref{model2})}

 The exact solution of model (\ref{model2}) was obtained in  \cite{AndJ}. 
The spectrum includes heavy solitons and antisolitons  with mass $M$ and a light singlet Majorana fermion with mass $m$. The mass ratio is $m/M \sim (\lambda - \lambda')$ and vanishes in the limit of equal couplings. This is the limit which was studied in \cite{NTsv}, it corresponds to the uniform interchain interactions and periodic boundary conditions in the transverse direction. The presence of such periodicity constitutes an   additional symmetry  whose  presence leads to vanishing of certain correlation functions. Therefore we prefer to keep $m/M$ finite. The two-particle scattering matrix of model (\ref{model2}) is given by  
\bea
S(\theta) = \left(
\begin{array}{lr}
S[su(2);\theta]_{\s_1,\s_2}^{\bar\s_1,\bar\s_2} & \delta_{\s,\bar \s}
\frac{\re^{\theta/2} - \ri}{\re^{\theta/2} + \ri}\\
\delta_{\s,\bar \s}\frac{\re^{\theta/2} - \ri}{\re^{\theta/2} + \ri} & -1
\end{array}
\right)
\label{S}
\eea
where $S[su(2)]$ is a 2$\times$2 scattering matrix of 
the SU(2) Thirring model.

The Thermodynamic Bethe Ansatz (TBA) 
are
\bea
F/L = - \frac{Tm}{2\pi}\int \rd\theta \cosh\theta\ln[1 + \re^{\epsilon_1(\theta)/T}] - \frac{TM}{2\pi}\int \rd\theta \cosh\theta\ln[1 + \re^{\epsilon_2(\theta)/T}]
\eea
\bea
\epsilon_n(\theta) &=& Ts*\ln[1 + \re^{\epsilon_{n -1}(\theta)/T}][1 + \re^{\epsilon_{n +1}(\theta)/T}] \nonumber\\
&-& \delta_{n,1}m\cosh\theta - \delta_{n,2}M\cosh\theta
\eea
These equations are valid for $\lambda' \approx \lambda$. 

 It is worth mentioning that  model (\ref{model2}) is generated  in the relativistic limit as  the spin sector of the following fermionic model: 
\bea
&&H = \int \rd x\left[\sum_{j = 1}^2 C_{j\s}^+(- \p_x^2 - k_F^2)
C_{j\s} + t(C^+_{1\s}C_{2\s} + C^+_{2\s}C_{1\s}) - gC^+_{j\s}C_{k\s}C^+_{k\s'}C_{j\s'}\right] , \nonumber\\
&&\s = \pm 1/2, j = 1,2; ~~ g >0 \label{model1}
\eea
From the solution of this model obtained in  \cite{tsvelik} it can be extracted that 
\bea
&&\frac{m}{M} = \frac{\pi}{8}\left(\frac{t_{\perp}}{\epsilon_F}\right)^2\ln^2(\epsilon_F/M)\nonumber\\
&&M = \frac{4}{\pi^2}k_F g\exp(- \pi k_F/g)
\eea
which provides a direct relationship with  model (\ref{model2}). 


\section{The correlation functions}

 We shall calculate the correlation functions using the formfactor approach. In this approach one calculates  matrix elements (formfactors) of various operators using their transformation properties under various symmetries of the problem. General information about the method can be obtained from various review articles \cite{fedya},\cite{grach}. 

Let us consider operators $\exp[\ri\s\sqrt{2\pi}\varphi_1]$ and $\exp[\ri\s\sqrt{2\pi}\varphi_3]$. We will be  interested only in the limit $m << M$, therefore it is  sufficient to consider the states with one soliton (antisoliton) and arbitrary number of light particles. In this case it is reasonable to suggest the following formfactor expansions:
\bea
&&\exp[\ri\s\sqrt{2\pi}\varphi_1]|0> = Q_{even} + Q_{odd}\nonumber\\
&&\exp[\ri\s\sqrt{2\pi}\varphi_3]|0> = Q_{even} - Q_{odd} \label{exp}
\eea
where $Q_{even}$ ($Q_{odd}$) has  matrix elements with the states with of one heavy 
soliton and even (odd) number of light Majorana fermions. Notice that the on-chain correlation functions for both operators are the same as it must be. The problem is simplified by the fact that the Majorana fermions have a non-trivial scattering matrix only with the heavy particle. Otherwise the dependence of the matrix element on the Majorana fermion rapidities is like for the Ising model. Using the results of \cite{Smirnov}, where a systems with a similar S-matrix was studied, we obtain for the following expressions:
\bea 
&&<s(\beta);\chi_1(\beta_1),...\chi_{2n}(\beta_2n)|Q_{even} = (C^+)^{1/2}\re^{\beta/4}\prod_{j=1}^{2n}\psi^{(-)}(\beta - \beta_j)\prod_{i.j}\tanh\left(\beta_{ij}/2\right)\label{form1}\\
&&<s(\beta);\chi_1(\beta_1),...\chi_{2n +1}(\beta_{2n+1})|Q_{odd} = {C^-}^{1/2}\re^{\beta/4}\prod_{j=1}^{2n+1}\psi^{(-)}(\beta - \beta_j)\prod_{i.j}\tanh\left(\beta_{ij}/2\right) \label{form2}
\eea
where $\beta$ is  the soliton's rapidity and $\beta_i$ stand for the rapidities of the light particles. $C^{\pm}$ are  normalization factors to be determined later. The function $\psi^{(-)}(\beta)$ satisfies the following equation
\bea
\psi^{(-)}(\beta)\psi^{(-)}(\beta - \ri\pi) = \frac{1}{1 + \ri\re^{\beta}} \label{psi}
\eea
and is given by 
\bea
&&\psi^{(-)}(\beta) = \re^{-\beta/4}\psi^{(0)}(\beta), \\
&&\psi^{(0)}(\beta) = 2^{-3/4}\exp\left\{ - \int_0^{\infty}\frac{\rd\omega}{\omega} \frac{2\sin^2[(\beta + \ri\pi)\omega/2] + \sinh^2(\pi\omega/2)}{2\sinh(\pi\omega)\cosh(\pi\omega/2)}\right\} \nonumber
\eea
As follows from (\ref{psi}), this function has the following asymptotics:
\bea
&&\psi^{(-)}(\beta \rightarrow +\infty) \rightarrow \ri\re^{- \beta/2}\nonumber\\
&&\psi^{(-)}(\beta \rightarrow -\infty) \rightarrow  1
\eea
The operators $\exp[\ri\sqrt{2\pi}\bar\varphi(2,4)]$ are expressed in a similar way, but with all rapidities having the opposite sign. Therefore their 
expansions include the function $\psi^{(+)}(\beta) = \psi^{(-)}(-\beta)$ which decays at $- \infty$ and approaches 1 at $\beta \rightarrow +\infty$.  

 Let us briefly comment on expressions (\ref{form1},\ref{form2}). The operators  under consideration have Lorentz spin 1/4. This dictates that under the Lorentz transformation $\beta \rightarrow \beta + \theta, \beta_j \rightarrow \beta_j + \theta$ the  matrix element of that operator must acquire a factor $\exp(\theta/4)$. As we see from (\ref{form1},\ref{form2}), this property is fulfilled. Another property is that the formfactor is multiplied on the S-matrix $S(\beta,\beta')$ whenever two particles with rapidities $\beta$ and $\beta'$ are interchanged. Since $\tanh(\beta_{ij}/2)$ is an odd function, this is obviously fulfilled for the Majorana fermions. The function $\psi$ plays the same role for the ineterchange of a Majorana fermion and the heavy particle. The choice of asymptotics of functions $\psi^{(\pm)}$ is determined by the fact that in on small distances bosonic exponents (\ref{exp}) become (anti)holomorphic fields. 
      
 Substituting (\ref{form1},\ref{form2}) into Eqs.(\ref{exp}) and performing certain algebraic manipulations (see Appendix 1), we arrive at the following expressions for the correlation functions on the same and on different chains ($r^2 = \tau^2 + x^2$, we put $v =1$):
\bea
&&G_{pq} \equiv \la\la \re^{\ri\sqrt{2\pi}\varphi(p)}\re^{-\ri\sqrt{2\pi}\varphi(q)} \ra\ra = \left(\frac{\tau - \ri x}{\tau + \ri x}\right)^{1/4}C(M^4m)^{1/8}{\cal F}_{pq}(r)\nonumber\\
&&D_{pq} \equiv \la\la \re^{\ri\sqrt{2\pi}\varphi(p)}\re^{\ri\sqrt{2\pi}\bar\varphi(q)} \ra\ra = C(M^4m)^{1/8}{\cal F}_{pq}(r) \label{GF}
\eea
where $C$ is a numerical coefficient, $p,q = 1,3$ for $G$ and $p=1,3; q = 2,4$ for $D$. 
\bea
&&{\cal F}_{11}(r) = {\cal F}_{33}(r) = \nonumber\\
&& = \int\rd\beta \re^{\beta/2}\re^{- Mr\cosh\beta}\exp\left\{\sum_{n=1}\frac{1}{n}\int \rd\beta\prod_{j}^n|\psi^{(-)}(\beta - \beta_j)|^2\re^{-mr\cosh\beta_j}\prod_{i>j}[\cosh(\beta_{ij}/2)]^{-1}\right\}\\
&&{\cal F}_{13}(r) = {\cal F}_{24}(r) = \nonumber\\
&& = \int\rd\beta \re^{\beta/2}\re^{- Mr\cosh\beta}\exp\left\{\sum_{n=1}\frac{(-1)^n}{n}\int \rd\beta\prod_{j}^n|\psi^{(-)}(\beta - \beta_j)|^2\re^{-mr\cosh\beta_j}\prod_{i>j}[\cosh(\beta_{ij}/2)]^{-1}\right\}
\eea
For the correlation functions on  neighboring chains (see Eqs.(\ref{corr})
($p =1,3; q = 2,4$) we  have 
\bea
&&D_{12}(r) = \int\rd\beta \re^{- Mr\cosh\beta}\exp\left\{\sum_{n=1}\frac{1}{n}\int \rd\beta\prod_{j}^n|\psi^{(0)}(\beta - \beta_j)|^2\re^{-mr\cosh\beta_j}\prod_{i>j}[\cosh(\beta_{ij}/2)]^{-1}\right\}\nonumber\\
&&D_{14}(r) = \int\rd\beta \re^{- Mr\cosh\beta}\exp\left\{\sum_{n=1}\frac{(-1)^n}{n}\int \rd\beta\prod_{j}^n|\psi^{(0)}(\beta - \beta_j)|^2\re^{-mr\cosh\beta_j}\prod_{i>j}[\cosh(\beta_{ij}/2)]^{-1}\right\}\label{D}
\eea

 The normalization factor $(M^3m)^{1/8}$ in Eqs.(\ref{GF}) is dictated by the facts that (i) the single chain correlation functions do not vanish in the limit $m \rightarrow 0$, (ii) the scaling dimension of the operator is 1/4. In the limit $Mr >> 1, mr << 1$ the integral in $\beta$ converges at $|\beta| \sim (Mr)^{-1/2}$. Since $|\psi^{(0)}(\beta_j)|^2 \sim \exp(- |\beta_j|/2)$, the integrals in $\beta_j$ converge even at $mr =0$. As result we get
\bea
D_{12}(r) = C(M^3m)^{1/8}K_0(Mr), ~~ D_{14}(r) = C(M^3m)^{1/8}K_0(Mr)
\eea
where is $C$ the  numerical constant which remains undertermined. 
To calculate the other  asymptotics is a more complicated task. The calculations are  done in Appendix 2; the result is 
\bea
&&{\cal F}_{11}(r) = C_{11}M^{-1/4}\re^{-Mr}r^{-5/8}\nonumber\\
&&{\cal F}_{13}(r) = C_{13}m^{1/2}M^{-3/4}r^{-1/8}\re^{-Mr}
\eea
where $C$ are again undertermined numerical constants. From here we derive the following estimates for the singularities in the imaginary part of the staggered magnetic susceptibility (dimerization) at $s =2M, (s^2= \omega^2 - q^2)$:
\bea
\Im m\chi_{11}(s) \sim (s -2M)^{-1/4}, ~~\Im m\chi_{12} \sim - m^{1/4}(s -2M)^{-1/2}, ~~ \Im m\chi_{13}(s) \sim mM^{-5/4}\theta(s - 2M)
\eea 
These expressions are valid at $s - 2M >> m$ so that $\Im m\chi_{12}$ cannot really become greater than $\Im m\chi_{11}$.
  
\section{Conclusions}
  Looking at expressions for the interchain correlation functions, we see that thay are proportional to powers of the particle masses. These masses are generated dynamically in the theory, they are exponentially small in the inverse coupling constant and therefore the interchain tunneling of the solitons is an essentially non-perturbative process. This conclusion coincides with the result for two chains \cite{NTsv}. The four chain case, however, introduces a new feature: the presence of the singlet particle. It is interesting that the prefactor in the interchain correlation functions contains both masses and vanishes when the mass of the singlet particle goes to zero (this occurs for periodic boundary conditions in the transverse direction; the case considered in \cite{NTsv}). This indicates that the singlet sector plays an important role in the interchain propagation of solitons probably providing  conditions for an uninhibited tunneling. We remind the reader that the fraction of the Hilbert space occupied by singlet excitations  grows with the number of chains \cite{NTsv} and these  excitations will certainly play an important  role in the thermodynamic limit determining interchain soliton propagation.    

 In conclusion we would like to emphasise a curious symmetry between correlations functions of the staggered energy density and magnetization (\ref{corr}) existing in Confederate Flag magnet. This symmetry unites $\epsilon$ and ${\bf N}$ into a four-component vector giving rise to an analogy between this model and models of non-collinear magnets. As it was pointed out in \cite{chubukov}, spinons naturally exist in disordered non-collinear magnets and it looks likely that Confederate Flag model provides a microscopic  realization of that scenario.   

\section{Acknowledgements}

 We are  grateful to Alexander Its for valueable discussions and references.  AMT acknowledges the support from 
 US DOE under contract number DE-AC02 -98 CH 10886. FS acknowledges the support from the Institute for Strongly Correlated and Complex Systems at BNL and from INTAS grant number 00-00055.  

$
~~
$

{\bf APPENDIX 1}
 
\bea
&&G_{11} = \sum_{n=0}\frac{1}{n!}\int\rd\beta\re^{\beta/2}\re^{- M\tau\cosh\beta - \ri Mx\sinh\beta}\prod_{j}^n\rd\beta_j\prod_{j=1}^n|\psi^{(-)}(\beta - \beta_j)|^2\re^{- m\tau\cosh\beta_j - \ri xm\sinh\beta_j}\prod_{i<j}\tanh^2(\beta_{ij}/2)\nonumber\\
&&G_{13} = \sum_{n=0}\frac{(-1)^n}{n!}\int\rd\beta\re^{\beta/2}\re^{- M\tau\cosh\beta - \ri Mx\sinh\beta}\prod_{j}^n\rd\beta_j\prod_{j=1}^n|\psi^{(-)}(\beta - \beta_j)|^2\re^{- m\tau\cosh\beta_j - \ri xm\sinh\beta_j}\prod_{i<j}\tanh^2(\beta_{ij}/2)\nonumber\\
\eea
By the  shift of the integration contours one obtains the following expressions:
\bea
&&G_{pq}(\tau,x) = \left(\frac{\tau - \ri x}{\tau + \ri x}\right)^{1/4}{\cal F}_{pq}(r), ~~ r^2 = \tau^2 + x^2\\
&&{\cal F}_{11}(r) = {\cal F}_{33}(r) = \nonumber\\
&&\sum_{n=0}\frac{1}{n!}\int\rd\beta\prod_{j}^n\rd\beta_j\re^{\beta/2}\re^{- Mr\cosh\beta}\prod_{j=1}^n|\psi^{(-)}(\beta - \beta_j)|^2\re^{- mr\cosh\beta_j}\prod_{i<j}\tanh^2(\beta_{ij}/2)\nonumber\\
&& = \int\rd\beta \re^{\beta/2}\re^{- Mr\cosh\beta}\exp\left\{\sum_{n=1}\frac{1}{n}\int \rd\beta\prod_{j}^n|\psi^{(-)}(\beta - \beta_j)|^2\re^{-mr\cosh\beta_j}\prod_{i>j}[\cosh(\beta_{ij}/2)]^{-1}\right\}\\
&&{\cal F}_{13}(r) = {\cal F}_{31}(r) = \nonumber\\
&&\sum_{n=0}\frac{(-1)^n}{n!}\int\rd\beta\prod_{j}^n\rd\beta_j\re^{\beta/2}\re^{- Mr\cosh\beta}\prod_{j=1}^n|\psi^{(-)}(\beta - \beta_j)|^2\re^{- mr\cosh\beta_j}\prod_{i<j}\tanh^2(\beta_{ij}/2)\nonumber\\
&& = \int\rd\beta \re^{\beta/2}\re^{- Mr\cosh\beta}\exp\left\{\sum_{n=1}\frac{(-1)^n}{n}\int \rd\beta\prod_{j}^n|\psi^{(-)}(\beta - \beta_j)|^2\re^{-mr\cosh\beta_j}\prod_{i>j}[\cosh(\beta_{ij}/2)]^{-1}\right\}
\eea
Here we have used the fact that the correlation functions are equal to the determinants of the Fredholm operators (see Appendix 2) and used Eq.(\ref{trlog}).  
The correlation functions on neighboring chains are given by  similar integrals:
\bea
&&D_{pq} \equiv \la\la \re^{\ri\sqrt{2\pi}\varphi(1)}\re^{\ri\sqrt{2\pi}\bar\varphi(2)} \ra\ra = {\cal F}_{pq}(r)\\
&&{\cal F}_{12}(r) = \sum_{n=0}\frac{1}{n!}\int\rd\beta\prod_{j}^n\rd\beta_j\re^{- Mr\cosh\beta}\prod_{j=1}^n\psi^{(-)}(\beta - \beta_j)\psi^{(+)}
(\beta - \beta_j)\re^{- mr\cosh\beta_j}\prod_{i<j}\tanh^2(\beta_{ij}/2)\nonumber\\
&&{\cal F}_{14}(r) = \sum_{n=0}\frac{(-1)^n}{n!}\int\rd\beta\prod_{j}^n\rd\beta_j\re^{- Mr\cosh\beta}\prod_{j=1}^n\psi^{(-)}(\beta - \beta_j)\psi^{(+)}(\beta - \beta_j)\re^{- mr\cosh\beta_j}\prod_{i<j}\tanh^2(\beta_{ij}/2)\nonumber
\eea
Taking into account that  $\psi^{+}(\beta)\psi^{(-)}(\beta) = |\psi^{(0)}(\beta)|^2$ and using Eq.(\ref{trlog}) we arrive at Eqs.(\ref{D}). 

$
~~
$

{\bf APPENDIX 2}

In this Appendix we calculate the asymptotics of correlation functions ${\cal F}_{pq}(r)$.
First, let us consider as an example the scaling of the Ising model. It is well-known that the 
correlation functions of the order and disorder parameter operators for this model can be expressed as follows:
\begin{align}
& \langle \mu (x)\mu (0)\rangle =G_+(mr)+G_-(mr),\non\\
&\langle \sigma (x)\sigma (0)\rangle =G_+(mr)-G_-(mr),\non
\end{align}
\begin{align}
G_{\pm}(\rho)=\sum _{n=0}^{\infty}(\pm)^n\frac 1 {n!}
\int\limits _{-\infty}^{\infty}\rd\beta _1
\cdots \int\limits _{-\infty}^{\infty}\rd\beta _n  \re^{-\rho\sum \cosh \b_j}
\prod\limits _{i<j}\tanh ^2\frac 1 2(\b _i -\b _j)\non
\end{align}
The functions  $G_{\pm}$ are
determinants of the Fredholm operators:
$$G_{\pm}(\rho)=\text{det}\(I\pm K\)$$
where $K$ is the integral operator with the kernel:
$$K(\b _1,\b _2)=\re^{-\frac 1 2\rho (\cosh \b_1+\cosh \b_2)}
\frac 1 {\cosh \frac 1 2(\b _1-\b _2)}$$
From here one can conclude that $G_{\pm}$ satisfy the Painlev\'e equation \cite{Jimbo}
and perform a rather detailed analysis of these functions.
The analysis of asymptotic behavior is a simplier task however and can be performed  as follows.
Using the formula
\be
\log (\text{det}A)=\text{Tr}(\log A)\label{trlog}
\ee
one obtains:
\begin{align}
&\log\(G_{\pm}(\rho)\)=\sum\limits _{n=1}^{\infty}(\pm)^n \frac 1 n
\int\limits _{-\infty}^{\infty}\rd\beta _1
\cdots \int\limits _{-\infty}^{\infty}\rd\beta _n  \re^{-\rho\sum \cosh \b_j}
\prod\limits _{i=1}^n\frac 1 {\cosh ^2 \frac 1 2 (\b_i-\b _{i+1})}\non
\end{align}
where $\b_{n+1}\equiv \b _1$.
Consider one of integrals:
\begin{align}
&\int\limits _{-\infty}^{\infty}\rd\beta _1
\cdots \int\limits _{-\infty}^{\infty}\rd\beta _n  \re^{-\rho\sum \cosh \b_j}
\prod\limits _{i=1}^n\frac 1 {\cosh ^2 \frac 1 2 (\b_i-\b _{i+1})}
\label{int}
\end{align}
We are interested in the 
asymptotics for $\rho\to 0$ where the integral diverges.
Let us introduce new variables:
\begin{align}
&\Delta _i=\b _i -\b _{i+1},\quad i=1,\cdots ,n-1,\non\\
&\omega =\sum\limits _{j=1}^N\cosh (\b _j)\label{change}
\end{align}
Then the integral becomes
\begin{align}
\int\limits _{-\infty}^{\infty}\rd\Delta _1
\cdots \int\limits _{-\infty}^{\infty}\rd\Delta _{n-1}
\ 2\int\limits _{D(\Delta)}^{\infty}\rd\omega\frac {\re^{-\rho\omega}}
{\sqrt{\omega ^2-D(\Delta)^2}}\prod\limits _{j=1}^{n-1}\frac 1 {\cosh\(\frac {\Delta _j}2\)}
\frac 1{\cosh\(\frac {\sum \Delta _j}2\)}\non
\end{align}
where 
$$D(\Delta _1,\cdots ,\Delta _{n-1})=\(\sum \re^{\b _j}\)\(\sum \re^{-\b _j}\)$$
It is clear that the integrals over  $\Delta _j$ are always rapidly converging
while the integral over  $\omega$ diverges at $\rho\to 0$. 
The estimation of leading contribution is straightforward. We give the 
final result for $G_+$ which diverges at $\rho\to 0$:
\begin{align}
&\log \(G_+(\rho)\)
\simeq -\[2\sum\limits _{j=1}^{\infty}
\int\limits _{-\infty}^{\infty}\rd\Delta _1
\cdots 
\int\limits _{-\infty}^{\infty}\rd\Delta _{n-1}
\prod\limits _{j=1}^{n-1}
\frac 1 {\cosh\(\frac {\Delta _j}2\)}
\frac 1 {\cosh\(\frac {\sum \Delta _j}2\)}
\]
\text{log} ( \rho)
\end{align}
The expression in square brackets is the anomalous dimension.
It  equals $\frac 1 4$. This value is known from
many sources, but actually even direct summation of series is possible. In a similar way we obtain
\bea
\log \(G_-(\rho)\) \sim \rho^{3/4}
\eea

Now let us turn to the case considered in the present paper. 
\begin{align}
&{\cal F}_{\pm}(\rho)=
\non\\
&=\sum _{n=0}^{\infty}(\pm)^n\frac 1 {n!}
\int\limits _{-\infty}^{\infty}\rd\beta _1
\cdots \int\limits _{-\infty}^{\infty}\rd\beta _n  \re^{-\rho\sum \cosh \b_j}
\prod\limits _{i<j}\tanh ^2\frac 1 2(\b _i -\b _j)
\prod\limits _{j=1}^n\left|\psi^{(-)}(\b _j)\right|^2\non
\end{align}
The functions ${\cal F}_{\pm}$ are determinants of the integral operators
$${\cal F}_{\pm}(\rho)=\text{det}\(I\pm L\)$$
where $L$ is the integral operator with the kernel:
$$L(\b _1,\b _2)=\re^{-\frac 1 2\rho (\cosh \b _1 +\cosh \b _2)}
\frac 1 {\cosh \frac 1 2(\b _1-\b _2)}
\left|\psi^{(-)}(\b _1)\psi^{(-)}(\b _2)\right|$$
So, similarly to the Ising case the logarithms of these determinants 
can be written as
\begin{align}
&\log\({\cal F}_{\pm}(\rho)\)=
\non\\&=\sum\limits _{n=1}^{\infty}(\pm)^n \frac 1 n
\int\limits _{-\infty}^{\infty}\rd\beta _1
\cdots \int\limits _{-\infty}^{\infty}\rd\beta _n  \re^{-\rho\sum \cosh \b_j}
\prod\limits _{i=1}^n\frac 1 {\cosh ^2 \frac 1 2 (\b_i-\b _{i+1})}
\left|\psi^{(-)}(\b_j)\right|^2\non
\end{align}
Consider the integral
$$
\int\limits _{-\infty}^{\infty}\rd\beta _1
\cdots \int\limits _{-\infty}^{\infty}\rd\beta _n  \re^{-\rho\sum \cosh \b_j}
\prod\limits _{i=1}^n\frac 1 {\cosh ^2 \frac 1 2 (\b_i-\b _{i+1})}
\left|\psi^{(-)}(\b _j)\right|^2
$$
and make  change of variables (\ref{change}). The difference with
the Ising case is due to the fact that we need to express $\b_j$ in terms
of the new variables and to substitute them into $\psi$. 
We have:
\begin{align}
&\b_j=\log\(\omega \mp \sqrt{\omega ^2-D^2(\Delta)}\)
-\log \(\re^{-\b_j}\sum\limits _k \re^{\b_k}\)
\label{b}
\end{align}
where the last term depends only on $\Delta$'s. The integral over $\omega$
is taken between the limits   $D(\Delta)$ and  $\infty$ over two branches:
one the  first one one takes  $+$ sign in (\ref{b}), and on the second one the minus sign. The part divergent at  $\rho\to 0$  comes from the region where $-\log (\omega)$ is large. It is not the second branch of the integral over $\omega$, since  $|\psi^{(-)}(\b)|$ is rapidly
decreasing as $\b\to+\infty$. So, only the first branch contributes,
and, since $|\psi^{(-)}(\b)|$ rapidly
 approaches  $1$ as $\b\to-\infty$, we can replace all $|\psi^{(-)}(\b_j)|$
by $1$. The result of this considerations is that the asymptotics
of the integral is $\frac 1 2$ of what we had in the Ising case. So, 
\begin{align}
\log \({\cal F}_+(\rho )\)\simeq -\frac 1 8 \log (\rho), ~~\log \({\cal F}_-(\rho )\)\simeq \frac 3 8 \log (\rho)\label{asym}
\end{align}

\end{document}